\documentstyle[11pt,mrs2001,epsfig]{article}
\def\void#1{{}}
\def\puncspace{\ifmmode\,\else{\ifcat.\C{\if.\C\else\if,\C\else\if?\C\else%
\if:\C\else\if;\C\else\if-\C\else\if)\C\else\if/\C\else\if]\C\else\if'\C%
\else\space\fi\fi\fi\fi\fi\fi\fi\fi\fi\fi}%
\else\if\empty\C\else\if\space\C\else\space\fi\fi\fi}\fi}
\def\SP{\let\\=\empty\futurelet\C\puncspace }
\def\etal{et\SP al.\SP }
\def\etal{{\it et al.\/}\ }

\def\lsim{~\rlap{$<$}{\lower 1.0ex\hbox{$\sim$}}}
\def\gsim{~\rlap{$>$}{\lower 1.0ex\hbox{$\sim$}}}
\def\iras{$IRAS$\SP}
\def\kms{kms$^{-1}$}
\def\hmpc{\,h{^{-1}}{\rm Mpc}}
\def\vvec{{\bf v}}
\def\rvec{{\bf r}}
\def\psc{{\it PSCz}\SP}
\def\iras{{\it IRAS}\SP}

\void{

\newcommand{\hmpc}{\ifmmode\,{\it h }^{-1}\,{\rm Mpc }\else $h^{-1}\,$Mpc\,\fi}

\newcommand{\hmpc}{\ifmmode\,{\it h }^{-1}\,{\rm Mpc }\else $h^{-1}\,$Mpc\,\fi}
\newcommand{\etal}{{\it et al.\ }}

\newcommand{\kms}{\ifmmode\,{\rm km}\,{\rm s}^{-1}\else km$\,$s$^{-1}$\fi} 

}

\begin{document}
\title{Matter in the Local Universe}

\author{L. da Costa} 
\affil{European Southern Observatory,
Karl-Schwarzschild Strasse 2, D-85748, Garching bei M\"unchen, Germany }

\begin{abstract}
In this contribution we review the large body of work carried out over
the past two decades to probe the dark matter in the local universe
using redshift survey and peculiar velocity data. While redshift
surveys have evolved rapidly over the years, gathering suitable
peculiar velocity data and understanding the short-comings of
different analyses have proven to be a difficult task. These
difficulties have led to conflicting results which have casted some
doubts on the usefulness of cosmic flows to constrain cosmological
models. Recently, however, a consistent picture seems to be finally
emerging with various methods of analyses applied to different data
sets yielding concordant results. These favor a low-density universe,
with constraints which are in good agreement with those obtained from
LSS, high-redshift supernovae and CMB studies.
\end{abstract}

\section{Introduction}

LSS studies of the nearby universe are arguably ideal to address the
question posed by the title of this conference. Indeed, if all the
mass in the universe were locked into galaxies, complete redshift
surveys of galaxies would provide the data required to fully
characterize the matter distribution. However, we have learned that
the luminous matter associated to galaxies represents a small fraction
of the mass density of the universe, and that galaxies may be biased
relative to the underlying distribution of matter. Still, if
structures grow as a result of gravity alone, observation of the
peculiar velocity of galaxies provides the means to probe the
distribution of the total matter. In the standard picture for the
formation of cosmic structures via gravitational instability the
peculiar velocity of a galaxy is generated by fluctuations in the mass
distribution. For galaxies outside virialized systems, linear
perturbation theory predicts
\begin{equation}
\vvec (\rvec) \approx {\Omega^{0.6}H_o \over 4\pi} \int{ d^3r^\prime \delta_m
{(\rvec^\prime - \rvec) \over \vert \rvec^\prime - \rvec \vert ^3}} \; .
\label{lingrav}
\end{equation}
This can also be expressed in the following differential form
\begin{equation} 
\nabla \cdot {\bf v} = -\Omega^{0.6} \delta_m,
\label{divv}
\end{equation}
where $\Omega$ is the mass density parameter, $H_o$ is the Hubble
constant and $\delta_m$ is the mass density fluctuation field. If
galaxies are fair tracers of the underlying mass distribution and
galaxy biasing is linear then $\delta_g = b \delta_m$, where
$\delta_g$ is the galaxy density contrast and $b$ is the bias
parameter for a given population of mass tracers.  The above equations
show that by mapping the peculiar velocity field one can determine the
distribution of mass and measure the parameter $\beta=\Omega^{0.6}/b$
by comparing the reconstructed density field with that observed for
galaxies or by comparing the measured velocity field with the
predicted gravity field generated by fluctuations of the galaxy
density field.

These simple ideas have been the underlying motivation for the major
wide-angle redshift surveys of optical and infrared galaxies and the
Tully-Fisher (TF) and $D_n-\sigma$ redshift-distance surveys conducted
over the past two decades.  In this contribution we review all of
these efforts, giving special emphasis to the results obtained from
recently completed redshift-distance surveys.  In section~\ref{z}, we
briefly mention the redshift surveys that have contributed to our
understanding of the local galaxy distribution and those which have
played a major role in the analysis of peculiar velocity data. In
Section~\ref{zv}, we review the redshift-distance surveys and the
peculiar velocity catalogs that have been used to map the peculiar
velocity field in the nearby universe. In section~\ref{results}, the
most recent surveys are used to reconstruct the velocity and density
fields and to measure $\beta$. These results are also compared with
those obtained in earlier works. Finally, in Section~\ref{summary} we
briefly summarize the current status of the field.

\section{Galaxy Distribution}
\label{z}

Over the past two decades the number of redshift surveys and redshift
data has greatly increased and a complete review is beyond the scope
of the present contribution and can be found
elsewhere~\cite{strauss-willick}\cite{dacostampaeso}. Here we point
out two classes of surveys that have had a strong bearing on some of
the issues discussed here. The first class consists of wide-angle,
dense sampling surveys such as the CfA2~\cite{gellerhuchra} and
SSRS2~\cite{dacostassrs} which revealed for the first time the full
complexity of the galaxy distribution. The discovery of extended,
coherent wall-like structures and of large regions devoid of luminous
matter with scales comparable to the survey depth represented a
serious challenge to the prevailing theories of structure formation
and evolution. Furthermore, these surveys probed relatively large
volumes which allowed for reasonable estimates of the power-spectrum
of the galaxy density fluctuations to be made for the first
time~\cite{gott}\cite{vogeley}\cite{dacostaps}. Even though unable to
reach very large scales, when COBE normalized, comparison with N-body
simulations demonstrated that the results were consistent with a
low-$\Omega$ cosmological model and an unbiased galaxy distribution.
The PS was well described by a shape parameter $\Gamma=0.2$,
consistent with other determinations~\cite{peakcockdodds}. Attempts
were also made to study the small-scale velocity field by analyzing
the redshift-space distortions. However, the small number of
independent structures within the sampled volume made the results
extremely sensitive to shot-noise~\cite{marzke}.

These early surveys were followed by the considerably deeper
LCRS~\cite{lcrs} which demonstrated unambiguously that the largest
scales of inhomogeneities had finally been reached. Quantitative
analyses of the LCRS, by and large confirmed earlier statistical
results, albeit with considerably smaller errors. More recently, the
first results of the 2dFGRS project have become available. The survey
consists of over 100,000 galaxies to a depth comparable to the LCRS,
allowing for precise measurements of redshift-space distortions and
large-scale power spectrum. Analysis of the redshift distortions
caused by large-scale infall velocities yields a value of
$\beta_o=0.43$~\cite{peakcock2df}, where $\beta_o$ refers to optical
galaxies. Assuming a relative bias $b_o/b_I\sim 1.3$ between optical
and \iras galaxies this implies $\beta_I\sim0.56$. The derived galaxy
power-spectrum~\cite{percival} was found to be well-represented by a
shape parameter $\Gamma=0.2$, in good agreement with previous
determinations. These results provide important constraints on the
mass power-spectrum which can later be compared with those obtained
using cosmic flows to test for consistency.

The second class of redshift surveys worth mentioning in the present
context is that involving galaxy samples extracted from the {\it IRAS}
survey such as the 1.9~Jy\cite{strauss1.9}, the 1.2~Jy\cite{fisher1.2}
and the \psc\ \cite{saunderspscz} redshift surveys.  While sparsely
sampling the galaxy distribution, these surveys provide a sky coverage
unmatched by optical surveys. This all-sky coverage allows a more
reliable determination of the gravity field induced by fluctuations of
the galaxy density field which can be compared to the measured
peculiar velocity field to estimate the parameter $\beta$.

From the above discussion, it is clear that by themselves redshift
surveys are more useful for studying the properties of galaxies than
as cosmological probes. However, combining them with redshift
independent distances to map out the peculiar velocity field of
galaxies and to predict the peculiar velocity field from galaxy
density fluctuations provide powerful tools to probe the nature of the
matter distribution and its relation with the galaxy distribution.

\section{Mapping the Peculiar Velocity Field}
\label{zv}

The radial component of the peculiar velocity is given by
$$U = cz - d$$ where $d$ is an estimate of the galaxy distance derived
from a secondary distance indicator. Most of the available samples
rely on the TF and Fundamental Plane relations for spirals and
early-type galaxies, respectively, with typical errors in distance of
$\sim 20\%$. However, samples based on distance indicators with
significant smaller errors, such as those based on surface
brightness fluctuations and nearby Type Ia supernovae, are slowly
growing and have already been successfully used to measure
$\beta$\cite{tonry}\cite{riess}.

In contrast to the rapid growth of samples with complete redshift
information, redshift-distance samples have been difficult to gather.
There are various practical reasons for that. First, to ensure the
uniformity of the data and of the sky coverage requires coordinated
observations in both hemispheres. Second, TF distances require the
measurement of the rotational velocity of the galaxy either from the
HI line width, which can only be efficiently measured in the northern
hemisphere, or from measurements of optical rotation curves, a
challenging observation. Third, for early-type galaxies, high
signal-to-noise spectra are required for accurate measurements of the
velocity dispersion. Finally, both distance indicators require
high-quality photometric data. Table~\ref{rd} summarizes the
redshift-distance surveys conducted to date. The table includes only
wide-angle redshift-distance surveys and the number of objects is just
indicative of the sample size. Not included are the various surveys
conducted to measure distances and peculiar velocities of clusters of
galaxies which have been used to constrain the amplitude of the bulk
flow on very large scales.

\begin{table}[t]
\caption{{\bf Summary of Redshift-Distance Surveys}}
\label{rd}
\begin{center}
\begin{tabular}{lrll}
\\
\hline
Survey  &    $ N_{obj}$  & Type & Coverage  \\
\hline \\
Aaronson \etal  & 300 &  spirals & all-sky \\
Tonry \& Davis   & 300  & early & north \\
7 Samurai        & 400  & early & all-sky  \\
Willick        & 320  & spirals   & Perseus-Pisces\\
Courteau \etal & 380 & Sb-Sc  & north \\
Mathewson \etal  & 2000+ & spirals & south \\
SFI              & 1300 & Sbc-Sc  & all-sky \\
ENEAR           & 1600  & early &  all-sky \\
Shellflow       & 300   & Sb-Sc & all-sky \\
\\
\hline\\
\end {tabular}
\end{center}
\end{table}

Early studies \cite{tonry-davis}\cite{aaronson} focused on the
properties of the flow field near Virgo. However, it was soon realized
that the assumption of a spherical infall was too restrictive and that
Virgo alone could not explain the motion of the Local Group relative
to the CMB. A major contribution to the field was the work of the
7~Samurai\cite{lynden-bell}, the first to probe well beyond the local
supercluster, albeit sparsely. Analyses of this sample led to
startling results such as the measurement of a large amplitude bulk
flow, and the discovery of the Great Attractor, a large mass
concentration associated with the Hydra-Centaurus complex. Among the
main conclusions of this work was that the large peculiar velocities
measured implied large values of $\Omega$, a result that placed cosmic
flows at odds with several other analyses.  By the end of the 80's the
first attempt to produce a homogeneous catalog by merging different
peculiar velocity data set was made (Mark~II) and used to obtain the
first map of the dark matter in the nearby universe~\cite{potent}.
Even though providing an important first glimpse of the dark matter
distribution, this early map showed that important regions of the sky
were severely undersampled.

In subsequent years a major effort was made to expand the available
sample to confirm the conclusions of the 7S and to improve the mass
maps. These efforts included small surveys of specific areas of the
sky~\cite{courteausurvey}\cite{willickdata} and major TF surveys of
spirals, such as those carried out by Matthewson and
collaborators~\cite{mat1}\cite{mat2} and the SFI
survey~\cite{haynes1}\cite{haynes2}, and FP surveys of early-type
galaxies, such as the recently completed ENEAR survey~\cite{enear}

Trying to capitalize as much as possible on all of the available data,
Willick and collaborators\cite{markiii} assembled the data from these
different surveys into a catalog (Mark~III) consisting of about 3000
galaxies, predominantly spirals, with measured peculiar velocities.
The Mark~III catalog, which does not include the more recent all-sky
SFI and ENEAR surveys, has been extensively used in the analyses of
peculiar velocity data. While considerable effort was made to ensure
uniformity, it is a compilation of heterogeneous data sets. As
illustrated in figure~11 of Kollat \etal\ \cite{kollat}, it lacks
uniformity in sky coverage due to the uneven coverage of the main data
sets included in the compilation. While the availability of this
catalog prompted the development of several techniques to analyze
peculiar velocity data and efforts to understand possible bias, its
use has led to conflicting results. The reasons for the discrepancies
are not understood and could indicate limitations of the data or of
the methods used. Efforts to re-calibrate this catalog using new
observations~\cite{shellflow} are still underway.

In this context the completion of the SFI I-band TF survey of late
spirals and the ENEAR $D_n-\sigma$ survey of early-type galaxies are
important additions, providing homogeneous samples of comparable
sizes. Figure~\ref{fig:skydist} shows the projected distribution of
galaxies in these two surveys.  In contrast to the Mark~III
compilation, the sky coverage of both surveys is remarkably uniform
and nearly all of the data consist of new measurements reduced in a
uniform way. Also note that the surveys nicely complement each other:
ENEAR galaxies probe high density regions and delineate large-scale
structures more sharply; SFI galaxies probe lower density regimes and
are more uniformly distributed across the sky.  Another important
point is that the peculiar velocities in these catalogs are measured
using distinct distance estimators based on different observed
quantities. Therefore, to take full advantage of these characteristics
these samples have been analyzed separately, to test the
reproducibility of the results, and combined into the SEcat catalog to
produce a fair sample probing a wide range of density regimes. The
results of analyses based on these new catalogs of peculiar velocity
data are reviewed below and compared to those obtained using Mark~III.

\begin{figure}
\caption{Projected distribution in Galactic coordinates of ENEAR (top)
and SFI(bottom) galaxies. }
\label{fig:skydist}
\end{figure}


\void{
It is convenient to distinguish between $b_I$ and $b_o$, the relative
of \iras and optical galaxies. Here we assume that
$\beta_o/\beta_I=0.75$ and that $b_o= 1/\sigma_8$, where $\sigma_8$ is
the $rms$ fluctuation amplitude within a sphere of $8~h^{-1}$~Mpc
radius. 
While adopting the same underlying assumptions different methods have
been developed to carry out velocity-velocity comparisons such as the
so-called ITF method\cite{nd} and VELMOD\cite{velmod}.
An alternative way of estimating $\beta$ or, more precisely the
parameter $\eta_8=\sigma_8\Omega^{0.6}$ is to apply a maximum
likelihood technique to the peculiar velocity data and derive the
best-fit model of the mass power spectrum, assuming the perturbations
to be in the linear regime\cite{wf}. One important advantage of the
method is that the derived PS is used as a prior for the
reconstruction of the velocity and density fields by means of the
Wiener filter technique.
}

\section{Results}
\label{results}

\void{
Over the past few years the methods described in previous section have
been applied to the several catalogs of peculiar velocity described
earlier. In this section the more recent results obtained using the
SFI, ENEAR and the combined SEcat catalogs are summarized and compared
with those obtained earlier using the Mark~III catalog. From this
comparison conclusions regarding the robustness and the
reproducibility of the results can be drawn.}

\subsection{Reconstructed density and velocity fields} 

An underlying assumption of all methods used in estimating $\beta$ is
that galaxies, even though biased, are fair tracers of the mass
distribution. This hypothesis can be, in principle, directly tested by
comparing the galaxy density field as derived from redshift surveys
and the mass density field reconstructed from peculiar velocity data.
POTENT, Wiener Filter\cite{wf} and more recently the Unbiased Minimal
Variance estimator (UMV)\cite{umv} are examples of methods developed
to reconstruct the three-dimensional velocity and density fields from
the observed radial component of the peculiar velocity.  All methods
assume that on the scales of interest the perturbations are small and
non-linear effects can be neglected. The various methods have also
been extensively tested using mock catalogs drawn from simulations
that mimic the nearby universe.

\begin{figure}
\plottwo{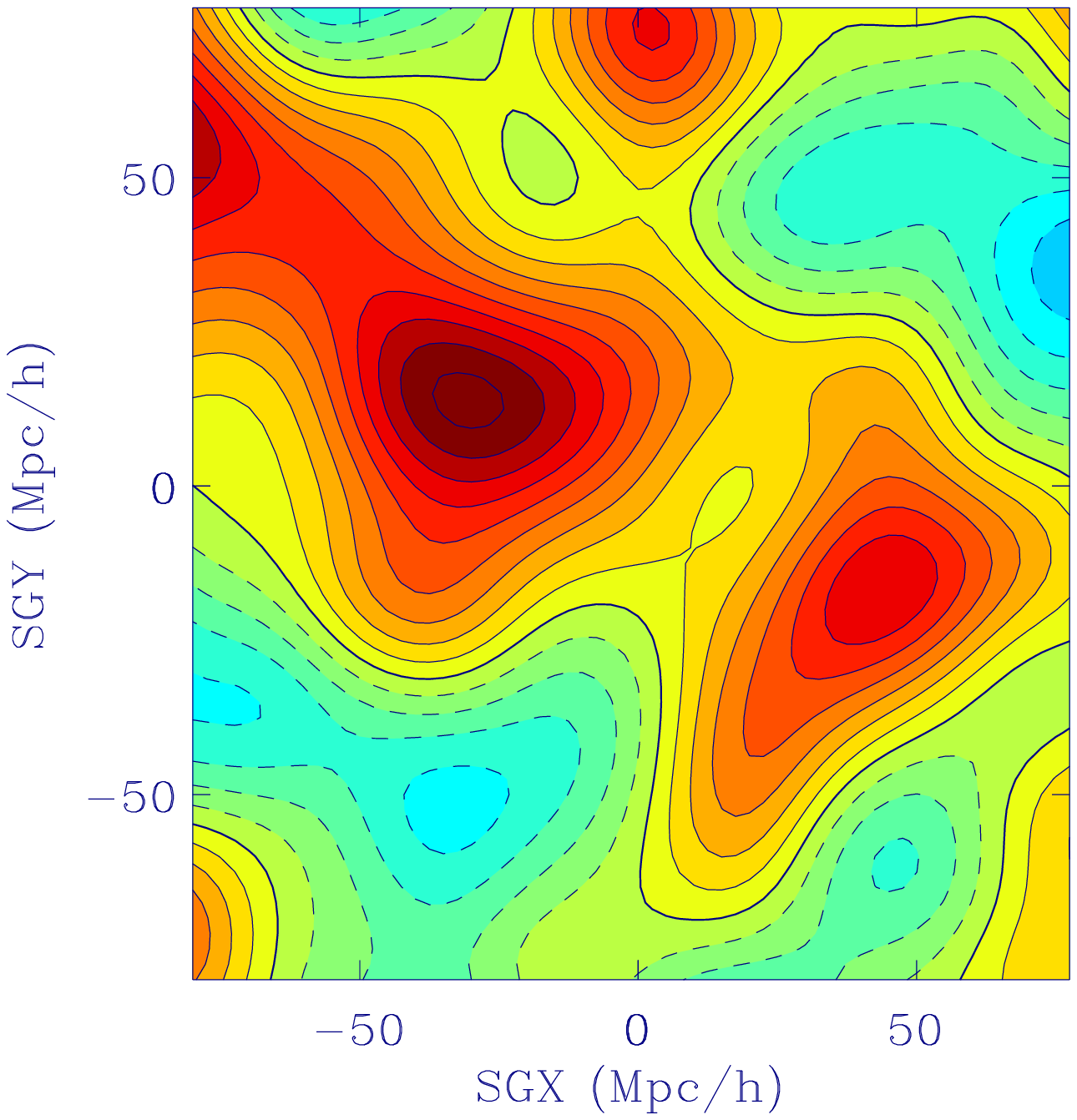}{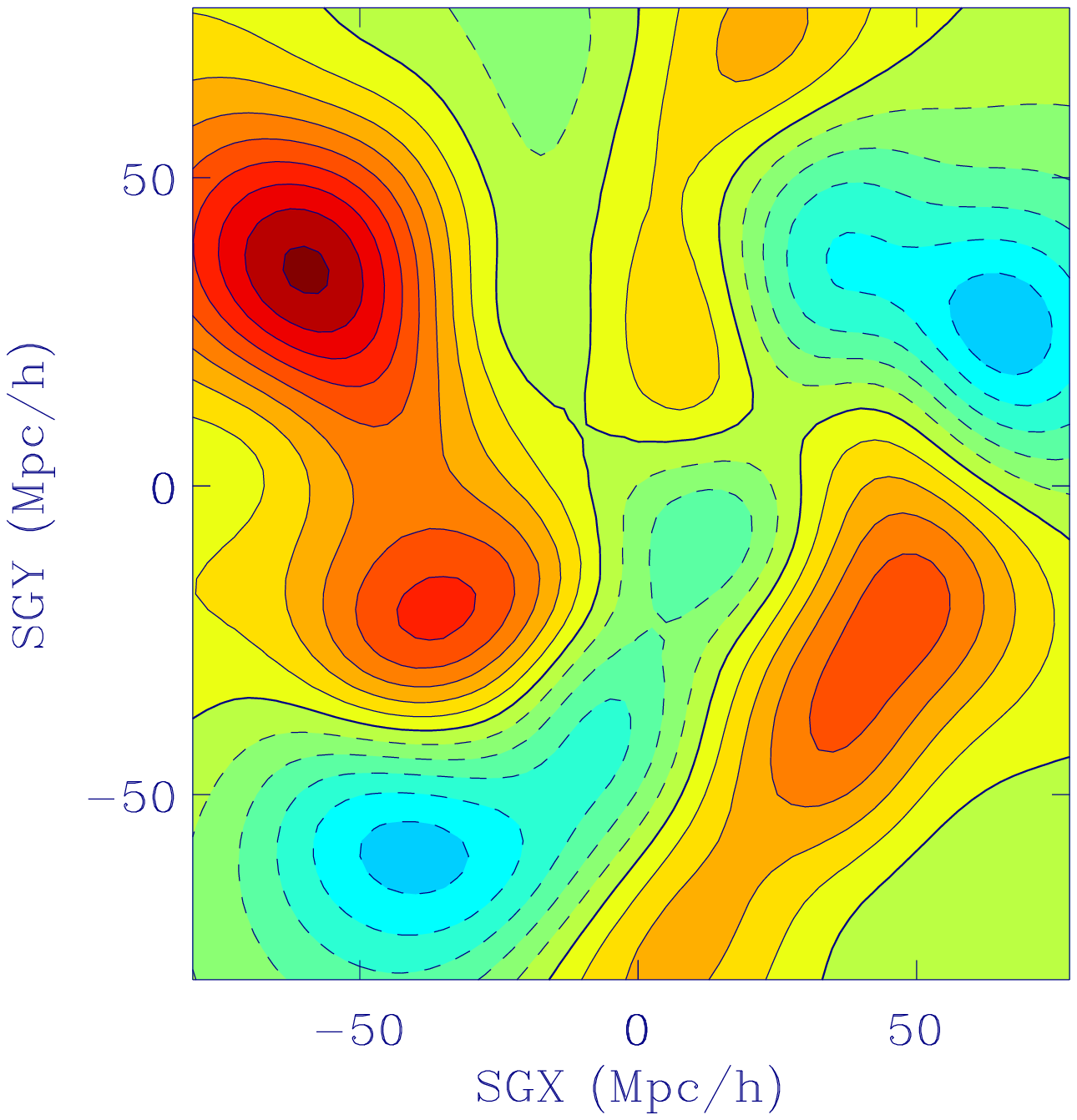}
\caption{ The reconstructed UMV Supergalactic plane density from the
{\it PSCz} (left panel) and the SEcat catalog (right panel) smoothed
with $12 h^{-1}\;$ Mpc Gaussian window. The solid and dashed line
contours denote positive and negative densities respectively. The
bold-solid line denote the zero level density. The contour spacing is
0.1. }
\label{fig:mass}
\end{figure}

Recently, the UMV method has been applied to the SEcat catalog of
peculiar velocities and to the \psc redshift survey data.
Figure~\ref{fig:mass} shows the map of the PSCz galaxy density field
(left panel) and the mass density field (right panel) along the
Supergalactic plane, the latter obtained from the SEcat data using a
Gaussian smoothing radius of 1200 \kms. The main features of our local
universe are easily identified in these maps, including the Great
Attractor (GA) on the left and the Perseus-Pisces supercluster (PP) in
the lower right. There is also a hint of the Coma cluster, which lies
just outside the volume probed by SEcat, in the upper part on the
map. The similarity between the mass and galaxy density fields is
striking, especially considering the limitations to the peculiar
velocity data imposed by the Zone of Avoidance. Furthermore, even
though different in details, the gross features of the mass density
field are similar to those obtained by applying either the same or the
POTENT formalism to the Mark~III catalog\cite{wfmark}\cite{potentm3}
and SFI catalogs\cite{dacostasc}. This is an outstanding result
considering the different ways these catalogs were constructed and the
peculiar velocities measured. Current results are also a remarkable
improvement over those obtained from earlier catalogs. In particular,
it is worth mentioning the prominence of the Perseus-Pisces region,
completely absent in the earlier maps, and the well-defined voids,
well-known features in redshift surveys which are now clearly seen in
the reconstructed mass distribution.

The reconstructed three-dimensional velocity field can also be used to
measure other quantities of interest. For instance, the amplitude of
the bulk flow is found to vary from $V_{B} =300 \pm$ 70~\kms\ for a
sphere of $R=20\hmpc$ to $160 \pm$ 60 \kms\ for $R=60\hmpc$.  This
value is in good agreement with that obtained from a direct fit to the
radial peculiar velocities for the SFI\cite{sdipole} and the
ENEAR\cite{edipole} samples.  This result disagrees with the bulk flow
determined for the Mark~III survey, which has an amplitude of roughly
twice this value\cite{wfmark}. The small amplitude of the bulk flow
recently measured is in marked contrast to earlier claims of large
amplitude coherent motions over scales of the order of
100$h^{-1}$~Mpc\cite{courteausurvey}, which at face value would imply
excess power on very large scales. This result is in line with the
results of recent redshift surveys which have not detected
inhomogeneities on very large scales.

Greater insight on the characteristic of the flowfield can be obtained
by decomposing the 3-D velocity field into two components, one which
is induced by the local mass distribution and a tidal component due to
mass fluctuations external to the volume
considered\cite{hoffman}\cite{wfmark}\cite{wfenear}.
Figure~\ref{fig:tidal} shows the results of this decomposition applied
to the ENEAR survey\cite{wfenear}, where the local volume is a sphere
of $ 80\hmpc$ centered on the Local Group.  The plots show the full
velocity field (upper left panel), the divergent (upper right panel)
and the tidal (lower left panel) components.  To further understand
the nature of the tidal field, its bulk velocity component has been
subtracted and the residual is shown in the lower right panel.  This
residual is clearly dominated by a quadrupole component.  In
principle, the analysis of this residual field can shed light on the
exterior mass distribution. For the ENEAR catalog we find that the
local dynamics is hardly affected by structure on scales larger than
its depth. For this sample not only the bulk velocity at large radii
is small but so is the $rms$ value of the tidal field, estimated to be
of the order of 60~\kms. This is in marked contrast to the results
obtained from the analysis of the Mark~III survey which yields a much
stronger tidal field, pointing (in the sense of its quadrupole moment)
towards the Shapley concentration. For Mark~III, the tidal field
contributes $\sim$ 200~\kms to the total bulk velocity.

\begin{figure}
\caption{Tidal field decomposition of the reconstructed velocity field
along the Supergalactic plane displayed as flow lines. The top-left
panel shows the full velocity field. The other panels are described in
the text.}
\label{fig:tidal}
\end{figure}

\subsection{Estimates of  $\beta$}

Equations (\ref{lingrav}) and (\ref{divv}) show that there are two
alternative ways for estimating $\beta$ - velocity-velocity or
density-density comparisons.  In the first case, the observed galaxy
distribution is used to infer a mass density field from which peculiar
velocities can be predicted and compared to the observed ones. In the
second case, the three-dimensional velocity field is obtained from the
observed radial velocities and used to infer a self-consistent mass
density field and thus a galaxy distribution, via linear biasing. The
latter is then compared to the one obtained from large all-sky
redshift surveys.

A particularly useful method for performing a velocity-velocity
comparison is the modal expansion method\cite{nd}. This method expands
the velocity fields by means of smooth functions (Bessel and spherical
harmonics) defined in redshift space, thus alleviating the Malmquist
biases inherent in real space analysis. Furthermore, the modal
expansion smooths the observed and predicted velocities in the same
way, so that the smoothed fields can be compared directly.  Because
the number of modes is substantially smaller than the number of data
points, the method also provides the means of estimating $\beta$ from
a likelihood analysis carried out on a mode-by-mode basis, instead of
galaxy-by-galaxy.  The similar smoothing and the mode-by-mode
comparison substantially simplify the error analysis.  The modal
expansion method has been used in comparisons between the \iras 1.2~Jy
predicted velocities and observed velocities inferred from TF
measurements\cite{dnw} in the Mark~III catalog, yielding $\beta_I\sim$
0.4. However, examination of the residual field showed a strong dipole
signature suggesting a significant mismatch between the Mark~III and
the \iras fields. The reasons for the mismatch are still not
well-understood.

\begin{figure}
\caption{Comparison of the smoothed velocity fields predicted from the
1.2~Jy {\it IRAS} survey (left) for an assumed value of $\beta_I=0.6$
and measured by the SFI redshift-distance survey (right).  The
velocity fields are shown in redshift shells 2000 kms$^{-1}$ thick.}
\label{fig:12sfi}
\end{figure}

More recently, the same method has been employed in the comparison of
the \iras 1.2~Jy and SFI\cite{dacostanusser} and of the \iras \psc and
ENEAR velocity fields\cite{nusserdacosta}. Figure~\ref{fig:12sfi}
shows the smoothed velocity field predicted from the 1.2~Jy \iras
survey (left), adopting the best-fit value of $\beta_I=0.6$, and the
measured SFI field (right). The infall to Virgo ($l =284^\circ, b =
74^\circ$) dominates the nearby SFI flow. In the middle panel, the
field exhibits a dipole pattern corresponding to the reflex motion of
the Local Group with infalling galaxies in the Hydra-Centaurus
direction and an outward flow in the Perseus-Pisces direction, as seen
in the LG restframe. Comparing the two fields one immediately sees
that the general pattern of the velocity fields is remarkably similar
with excellent agreement in the location of outflows and inflows and
with only a few nearby galaxies having large residuals. This result
gives confidence in the determination of $\beta_I$.  Most encouraging
is the absence of large regions of coherent residuals such as the
dipole signature seen in the Mark~III analysis at intermediate and
distant redshift shells. Similar analysis has been performed using the
\psc \iras survey and the ENEAR catalog of peculiar
velocities. Figure~\ref{fig:psczenear} shows the corresponding
smoothed velocity fields, for an adopted value of
$\beta_I=0.5$. Comparison of the right-side of Figures~\ref{fig:12sfi}
and \ref{fig:psczenear} shows that the general flow pattern of the SFI
and ENEAR velocity fields is remarkably similar. In the ENEAR case,
very few prominent structures are probed by bright ellipticals in the
innermost shell. However, in the next two shells a strong dipole
pattern can be easily recognized, having an amplitude comparable to
that observed in SFI. The agreement between the \psc and ENEAR
velocity fields is also very good with only a few more distant
galaxies having large residuals.

\begin{figure}
\caption{Comparison of the smoothed velocity fields predicted from the
{\it PSCz IRAS} survey (left), for an assumed value of $\beta_I=0.5$,
and measured by the ENEAR redshift-distance survey (right).  The
velocity fields are shown in redshift shells 2000 kms$^{-1}$ thick.}
\label{fig:psczenear}
\end{figure}

The above results demonstrate that the velocity fields of both SFI and
ENEAR are similar and well described by the gravity fields of the
\iras 1.2~Jy and \psc surveys, yielding comparable values of
$\beta_I$. Consistent values of $\beta_I$ have also been obtained from
similar analysis of the SBF survey of galaxy distances
($\beta_I=0.42$)\cite{sbfbeta} and from the peculiar velocities
measured for a sample of nearby Type Ia supernovae ($\beta_I=0.4$)
\cite{riess}. 

Another method to carry out a velocity-velocity comparison considered
is VELMOD, a maximum likelihood method which takes as input the
distance indicator observables and galaxy redshifts and determines the
parameters describing the distance relation and the velocity model
adopted. The method does not require smoothing and it is constructed
for high-resolution analysis. The method has been used to analyze
sub-samples of spiral galaxies extracted from the Mark~III
\cite{velmod}\cite{velmod1} and the SFI data\cite{branchini}, yielding
$\beta_I=0.49$ and $\beta_I=0.42$, respectively. These results show
that the value of $\beta_I$ obtained from velocity-velocity
comparisons is independent not only of the data set considered but
also of the method used, with all estimates being in the range
$0.4\lsim \beta_I \lsim 0.6$.

Unfortunately, until recently there has been a disparity between the
results obtained from velocity-velocity comparisons and other methods
such as density-density comparisons and maximum-likelihood estimates
of the power-spectrum (PS) of mass fluctuations derived from peculiar
velocity data \cite{zaroubips}. For instance, density-density
comparisons using different data sets have invariably led to high
values of $\beta$\cite{sigad}, consistent with unity. In particular,
comparison of the \iras 1.2~Jy and POTENT reconstructed density field,
based on the Mark~III catalog, yields $\beta_I=0.89$. Similarly,
estimates based on the PS derived from peculiar velocity data using
Mark~III\cite{zaroubips}, SFI\cite{freudling} and ENEAR\cite{wfenear}
have yielded values of $\beta$ in the range 0.82-1.1. The nature of
this discrepancy is unknown.  Both density-density and
velocity-velocity methods have been carefully tested using mock
catalogs extracted from N-body simulations and have been shown to
provide unbiased estimates of $\beta$. Possible reasons for the
discrepancy are non-linear effects, scale dependence of the biasing,
poorly understood errors and/or problems with the data. However,
attempts to evaluate their impact have so far failed to explain the
discrepancy. In general, velocity-velocity comparisons are considered
more robust as they depend more on redshift data, while
density-density comparisons uses less reliable peculiar velocity data.

\begin{figure}
\plottwo{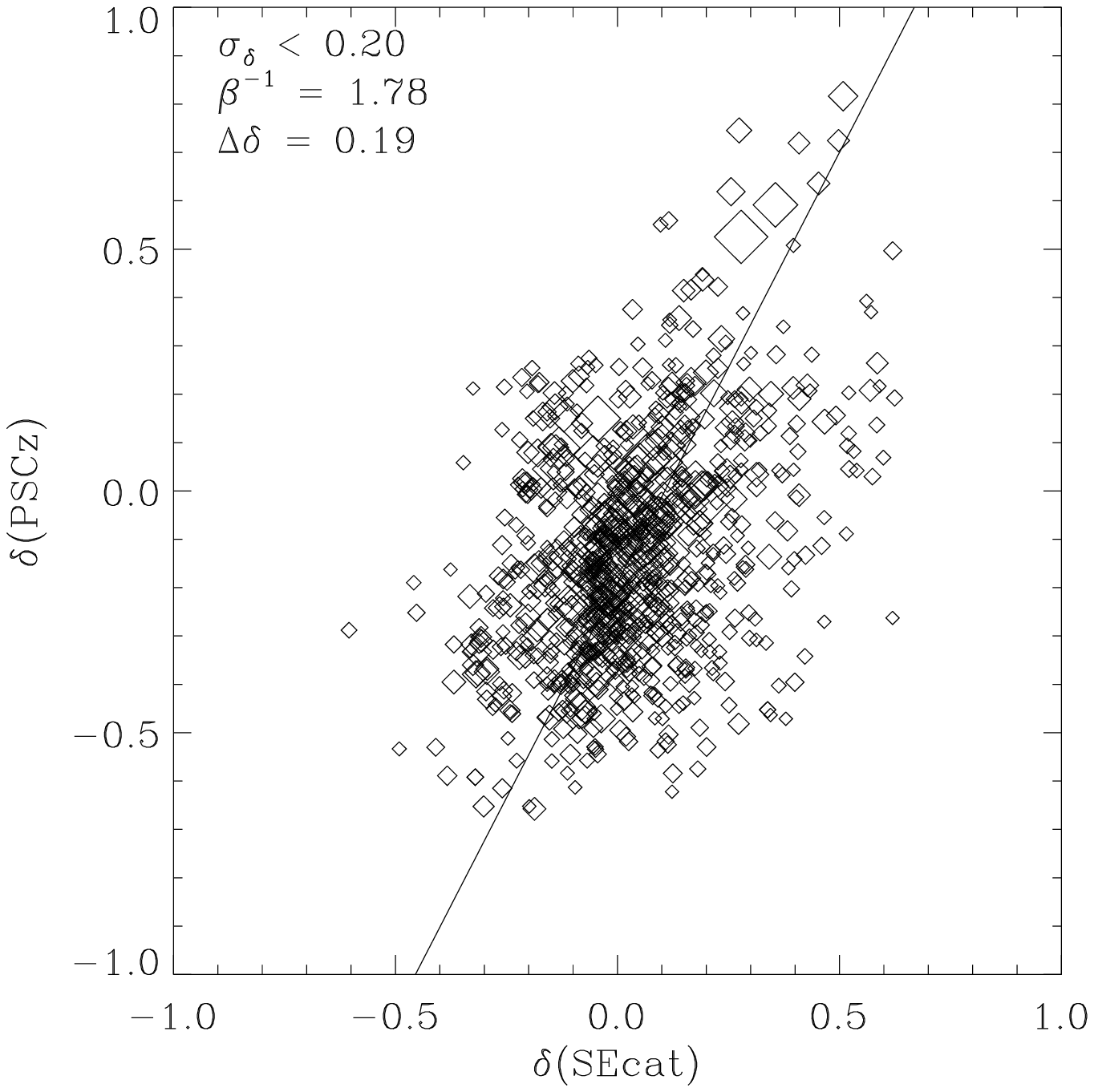}{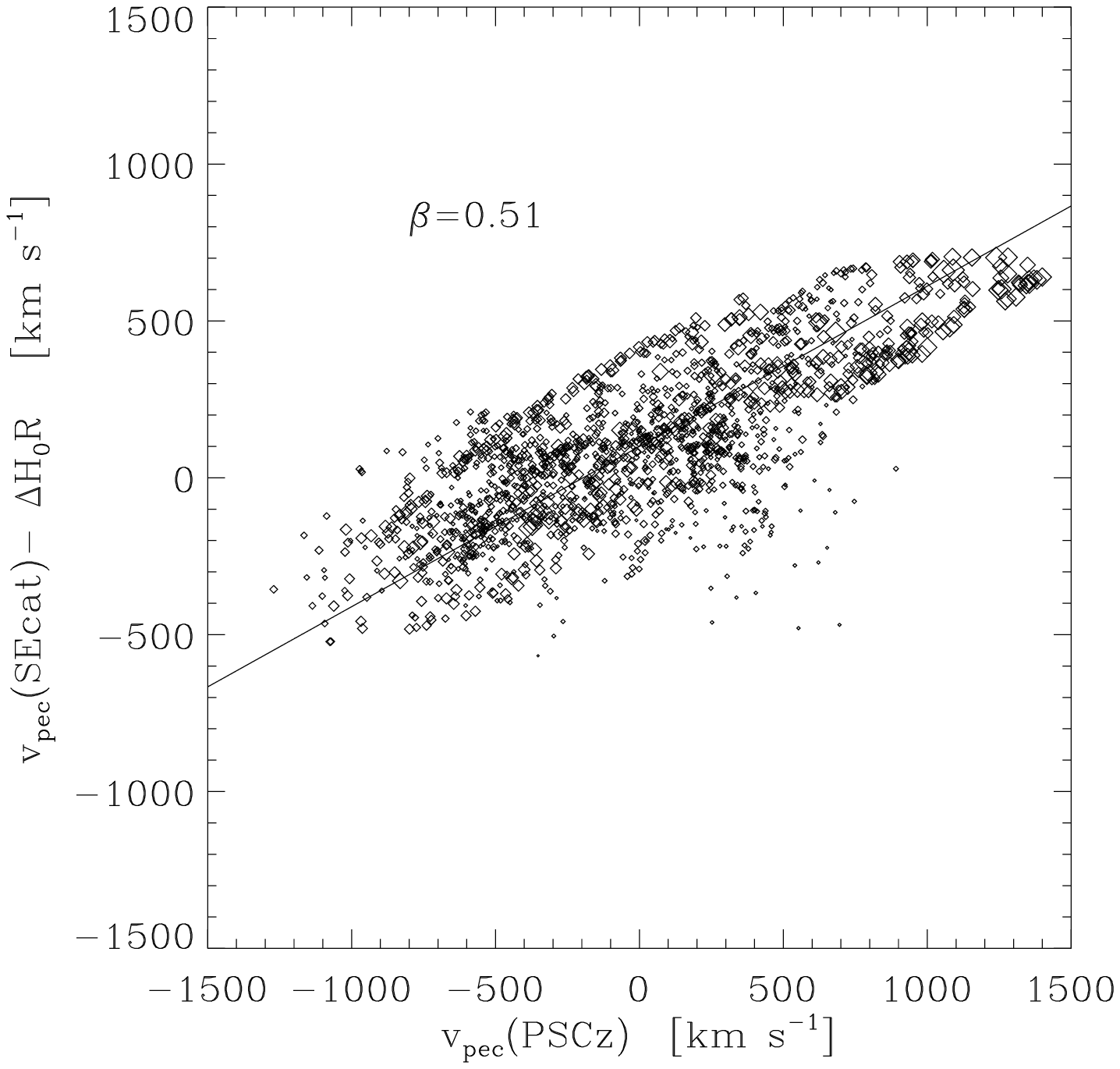}
\caption{Scatter plots showing the SEcat UMV reconstructed {\it vs.}
the PSC$z$ predicted densities (left panel) and velocities (right
panel). The size of the symbols is inversely proportional to the
errors}
\label{fig:umv}
\end{figure}

Recently, a new attempt to carry out a density-density comparison has
been made using the SEcat catalog mentioned earlier and the UMV method
to reconstruct the 3-D velocity and density fields. These
reconstructed fields were then used to determine the value of $\beta$
from direct velocity-velocity and density-density comparisons with the
corresponding fields predicted from the \psc redshift
survey\cite{secat}.  Figure~\ref{fig:umv} shows the results of the
density-density (left) and velocity-velocity (right) comparisons,
which give  $\beta_I=0.56 \pm 0.1$ and $\beta_I=0.51 \pm 0.05$,
respectively. This result is remarkable since it is the first time a
good agreement is found for $\beta$ values derived from these two
methods. This encouraging new result, which apparently resolves a
long-standing dispute, may be due either to the new method used in the
reconstruction of the fields or to the more homogeneous peculiar
velocity data used or a combination of both.

High values of $\beta$ have also been derived from applying a
maximum-likelihood technique to the peculiar velocity data to derive
the power-spectrum of mass density fluctuations. These results are
summarized in Figure~\ref{fig:ps}. In the left panel the PS obtained
from the ENEAR sample\cite{wfenear} with those measured for
Mark~III\cite{zaroubips} and SFI\cite{freudling}. The right panel
shows the contour map of the likelihood (in the $\Gamma-\eta_8$ plane)
for a $\Gamma$ model fit to the ENEAR data, where
$\eta_8=\sigma_8\Omega^{0.6}$ and $\sigma_8$ is the $rms$ fluctuation
amplitude within a sphere of $8~h^{-1}$~Mpc radius.  It is clear from
the figure that all data sets lead to similar high-amplitude PS,
equivalent to high values of $\beta$. From the figure one also can see
that while the likelihood analysis poses a strong constraint on
$\eta_8$, the value of $\Gamma$ is poorly determined. Note, however,
that low values of $\Gamma$, such as those required from the analyses
of redshift survey data ($\Gamma \sim0.2$), are excluded at about the
$3\sigma$ level.

\begin{figure}
\caption{Left panel: Power-spectrum of the most probable
COBE-normalized OCDM model estimated for the ENEAR data set compared
to those derived from Mark~III and SFI samples. Right panel: Contour
map of the likelihood for the $\Gamma$-model.}
\label{fig:ps}
\end{figure}

It is important to recall that the likelihood method used in
estimating the PS involves the use of model power-spectra to compute
the velocity correlation tensor which is then compared to that
computed from the peculiar velocity data to determine the fit
parameters. An equivalent way of exploring the same information is to
use the scalar velocity correlation function, computed under the
assumption of a homogeneous and isotropic flow. The results of this
analysis can then be compared directly to model predictions using
linear theory and an ensemble average of cosmic flow realizations for
different cosmological models. The statistics of the model velocity
field is parameterized by the amplitude, $\eta_8$, and by the shape
parameter, $\Gamma$, of a CDM--like power spectrum.  Applying the
velocity correlation statistics to the SFI\cite{borgani1} and
ENEAR\cite{borgani2} data sets one finds $\eta_8=0.34$ (SFI) and
$\eta_8=0.51$ (ENEAR) for $\Gamma=0.25$.  These values translate to
$\beta_I=0.45-0.67$, assuming $b_I/b_o\sim 1.3$, results which agree
within the uncertainties with the lower values of $\beta$ obtained by
other methods presented above. More importantly, in contrast to the PS
analyses, the region of acceptable solutions comfortably overlaps with
other constraints on $\eta_8$ derived from the $rms$ of cluster
peculiar velocities and cluster abundances, and on $\Gamma$ as
determined for the galaxy power spectrum.  One possible explanation
for the discrepancy between the results of the PS analysis and the
velocity correlation statistics is the different way the errors in the
distance measurements are taken into account. An important clue is the
weak constraint imposed on the shape parameter by the PS
analysis. This suggests that the available samples may not be
sufficiently deep for this type of analysis, making the method
insensitive to the effect that large scale power may have in inducing
velocities on small scales.

\section {Summary}
\label{summary}

After considerable effort on both the observational and theoretical
fronts, one can state with some degree of confidence that the most
controversial issues surrounding large-scale flows are being
resolved. The availability of different methods and of data sets have
enabled one to test the reproducibility of the results. Especially
important has been the completion of modern, homogeneous, all-sky
redshift-distance surveys of both spirals and early-type
galaxies. These samples probe comparable volumes and allow for
independent analyses. Contrary to earlier claims recent analyses yield
a small amplitude bulk flow, a mass distribution and velocity field
which closely resembles the galaxy density field and the associated
gravity field and concordant values of $\beta$ obtained using different
samples, distance indicators and methods. Current constraints argue in
favor of a low-density universe and are consistent with those set by
galaxy clustering, small-scale dynamics, present-day cluster
abundance, high-redshift supernovae and cosmic microwave
background. The agreement among such diverse measurements is not only
reassuring but gratifying for those who have worked so hard in the
field of cosmic flows.

\acknowledgements {I would like to thank all of my collaborators in
the SFI and ENEAR projects. Special thanks to S. Zaroubi for many
useful discussions.}

\end{document}